\documentclass[12pt]{article}
\usepackage[utf8]{inputenc}
\usepackage[british]{babel}
\usepackage{cmap}
\usepackage{lmodern}
\usepackage{float}
\usepackage{afterpage}
\usepackage[dvipsnames,svgnames,table]{xcolor}
\definecolor{fgcolor}{rgb}{0.345, 0.345, 0.345}
\newenvironment{knitrout}{}{} % an empty environment to be redefined in TeX

\usepackage{amssymb, amsmath, amsthm}
\usepackage[a4paper,top=25mm,bottom=25mm,left=25mm,right=25mm]{geometry}
\usepackage{ragged2e}

\usepackage{authblk} % for headings
\usepackage{pifont}
\usepackage{wasysym}
\usepackage{graphicx}
\usepackage[dvipsnames,svgnames,table]{xcolor}
\usepackage[figuresright]{rotating}
\usepackage{xtab} % tackle the long tables
\usepackage{longtable} % tackle the long tables
\usepackage{multirow}
\usepackage{footnote}
\usepackage[stable]{footmisc}
\usepackage{chngpage} % allows for temporary adjustment of side margins
\usepackage{pdflscape} % landscape environment
\usepackage[nottoc,notlot,notlof]{tocbibind} % includes everything in ToC

\usepackage{pgfplots}
\pgfplotsset{every tick label/.append style={font=\footnotesize}}
\pgfplotsset{compat=1.18}
\usepackage{setspace}

\usepackage{array}
\newcolumntype{K}[1]{>{\centering\arraybackslash$}p{#1}<{$}}

\makesavenoteenv{tabular}
\usepackage{tabularx}
\usepackage{booktabs}
\usepackage{threeparttable}
\usepackage[referable]{threeparttablex} % footnotes in tabu
\newcolumntype{R}{>{\raggedleft\arraybackslash}X}
\newcolumntype{L}{>{\raggedright\arraybackslash}X}
\newcolumntype{C}{>{\centering\arraybackslash}X}
\newcolumntype{A}{>{\columncolor{gray!25}}C}
\newcolumntype{a}{>{\columncolor{gray!25}}c}

\newlength{\tablen}

\usepackage{dcolumn} % alignment to decimal points
\newcolumntype{.}{D{.}{.}{-1}}

\usepackage{tikz}
\usetikzlibrary{arrows, calc, matrix, patterns, positioning, shapes, trees}
\usepackage[semicolon]{natbib} % numbers,
\usepackage[hyphens]{xurl}
\usepackage[nopatch=footnote]{microtype}
\usepackage[justification=centering]{caption} % [justification=centering]

\usepackage[labelformat=simple]{subcaption}

\DeclareCaptionLabelFormat{parenthesis}{(#2)}
\makeatletter
\renewcommand\p@subfigure{\arabic{figure}.}
\makeatother

\DeclareCaptionLabelFormat{parenthesis}{(#2)}
\makeatletter
\renewcommand\p@subtable{\arabic{table}.}
\makeatother

\usepackage{enumitem}

\setlist[itemize]{leftmargin=2.5\parindent}
\setlist[enumerate]{leftmargin=2.5\parindent}

\usepackage{hyperref} % [hidelinks]
\hypersetup{
  colorlinks   = true,    		% Colours links instead of ugly boxes
  urlcolor     = blue,    		% Colour for external hyperlinks
  linkcolor    = blue,    		% Colour of internal links
  citecolor    = ForestGreen	% Colour of citations
}

\def\addlegendimage{\csname pgfplots@addlegendimage\endcsname}

\theoremstyle{plain}
\newtheorem{claim}{Claim}%[section]

\theoremstyle{definition}
\newtheorem{definition}{Definition}%[section]
\newtheorem{example}{Example}%[section]

\theoremstyle{remark}

\newcommand{\up}{\textcolor{BrickRed}{\UParrow}}
\newcommand{\down}{\textcolor{PineGreen}{\DOWNarrow}}

\makeatletter
\let\@fnsymbol\@alph
\makeatother

\def\keywords{\vspace{.5em} % Add keywords
{\noindent \textit{Keywords}: }}

\def\AMS{\vspace{.5em} % Add keywords
{\noindent \textbf{\emph{MSC} class}: }}

\def\JEL{\vspace{.5em} % Add keywords
{\noindent \textbf{\emph{JEL} classification number}: }}

\def\Dedication{
{\noindent
``\emph{The following draw principles have been established to ensure a competitive balance -- both in the group stage and, depending on sporting results, into the knockout stage -- while retaining, insofar as possible, the random element inherent in tournament draws \citep{FIFA2025b}.}''
}
\vspace{1cm} 
\justify }

\title{On the non-uniformity of the \\ 2026 FIFA World Cup draw}

\author{
\href{https://sites.google.com/view/laszlocsato}{L\'aszl\'o Csat\'o}\thanks{~Corresponding author. Email: \emph{laszlo.csato@uni-corvinus.hu} \newline
Institute for Computer Science and Control (SZTAKI), Hungarian Research Network (HUN-REN), Laboratory on Engineering and Management Intelligence, Research Group of Operations Research and Decision Systems, Budapest, Hungary \newline
Corvinus University of Budapest (BCE), Institute of Operations and Decision Sciences, Department of Operations Research and Actuarial Sciences, Budapest, Hungary}
$\quad$
Martin Becker\thanks{~Email: \emph{martin.becker@mx.uni-saarland.de} \newline
Saarland University, Saarbr\"ucken, Germany}
$\quad$
Karel Devriesere\thanks{~Email: \emph{karel.devriesere@ugent.be} \newline
Ghent University, Department of Business Informatics and Operations Management, Ghent, Belgium \newline
FlandersMake@UGent -- core lab CVAMO, Ghent, Belgium}
$\quad$
Dries Goossens\thanks{~Email: \emph{dries.goossens@ugent.be} \newline
Ghent University, Department of Business Informatics and Operations Management, Ghent, Belgium \newline
FlandersMake@UGent -- core lab CVAMO, Ghent, Belgium}
}

\date{\today}

\begin{document}
\newgeometry{top=20mm,bottom=20mm,left=25mm,right=25mm}
\maketitle

\thispagestyle{empty}
\Dedication

\begin{abstract}
\noindent
%The group draw of a sports tournament can be used to increase attractiveness by extending the set of constraints required for a balanced draw. 
The group stage of a sports tournament is often made more appealing by introducing additional constraints in the group draw that promote an attractive and balanced group composition.
For example, the number of intra-regional group matches is minimised in several world championships. Under such constraints, the traditional draw procedure may become non-uniform, meaning that the feasible allocations of the teams into groups are not equally likely to occur.
Our paper quantifies this non-uniformity of the 2026 FIFA World Cup draw for the official draw procedure, as well as for 47 reasonable alternatives implied by all permutations of the four pots and two group labelling policies. We show why simulating with a recursive backtracking algorithm is intractable, and propose a workable implementation using integer programming. The official draw mechanism is found to be optimal based on two of five non-uniformity measures, including the average distortion of the 771 non-zero assignment probabilities. Nonetheless, non-uniformity can be more than halved if the organiser aims to treat the best teams drawn from the first pot equally.


\AMS{62-08, 90-10, 90B90, 91B14}
% Computational methods for problems pertaining to statistics
% Mathematical modeling or simulation for problems pertaining to operations research and mathematical programming
% Case-oriented studies in operations research
% Social Choice

\JEL{C44, C63, Z20}
% Operations Research, Statistical Decision Theory
% Computational Techniques, Simulation Modeling 
% Sports Economics, General
\end{abstract}

\clearpage
\restoregeometry

\section{Introduction} \label{Sec1}

The fundamental principle of a group stage draw in sports tournaments is to incorporate a random mechanism that treats all teams equally ex ante to the extent possible. In order to ensure balance, namely, to exclude groups of different overall strength ex post, usually additional draw constraints are designed, which imply equal treatment only for teams of comparable strength. To enhance the attractiveness of the tournament, further draw constraints can be imposed to guarantee that teams face opponents they would rarely encounter outside the tournament \citep{Csato2022d}. Draw constraints may also be required owing to security reasons \citep{Kobierecki2022}.
However, enforcing all draw constraints with a tractable and transparent random mechanism is far from straightforward \citep{BoczonWilson2023, Guyon2015a, RobertsRosenthal2024}. 

The governing body of (association) football, FIFA (F\'ed\'eration Internationale de Football Association), announced the rules of the 2026 FIFA World Cup draw on 25 November 2025. The draw took place on 5 December 2025. Although the regulation \citep{FIFA2025b} -- similar to the 2018 \citep{FIFA2017c} and 2022 \citep{FIFA2022a} editions -- does not specify how the draw constraints (see Section~\ref{Sec31}) are enforced, the video of the event  uncovers that the so-called \emph{Skip mechanism} (see Section~\ref{Sec32}) is used for this purpose.
% (\url{https://www.youtube.com/watch?v=9HX_tQBA-Iw})

According to \citet[p.~663]{RobertsRosenthal2024}, the Skip mechanism is ``\emph{surprisingly challenging to simulate with a computer program}''. Neither FIFA nor the Union of European Football Associations (UEFA), which employs a similar procedure called the \emph{Drop mechanism} in its competitions, provides an algorithm for this purpose. First in the literature, \citet{Guyon2014a} has suggested using backtracking to implement the Drop mechanism. Indeed, \citet{RobertsRosenthal2024} and \citet{Csato2025c} give recursive backtracking algorithms for the Skip mechanism that do work well for the 2018 and 2022 FIFA World Cups.

Our first main contribution is connected to simulating the 2026 FIFA World Cup draw. 
We show that, in contrast to previous editions of the tournament, a recursive backtracking algorithm remains hopelessly slow -- or requires some non-trivial and case-sensitive pre-filtering -- for the 2026 event due to the more restrictive constraints and the greater number of groups (see Section~\ref{Sec33}).
Therefore, a more efficient identification of deadlocks in advance is presented through an integer program, enabling a workable implementation of the Skip mechanism (see Section~\ref{Sec34}).
%As far as we know, our paper offers the first implementation of the Skip mechanism by an integer program. Although an integer programming approach has recently been suggested to simulate the UEFA Champions League league phase draw \citep{DevriesereGoossensSpieksma2026, GuyonBenSalemBuchholtzerTanre2026}, its draw uses the Drop mechanism, a procedure that differs from the Skip mechanism.
Both the Skip and Drop mechanisms are known to distort assignment probabilities compared to a uniform draw, which might benefit some teams at the expense of others \citep{Csato2025c, Guyon2014a, Kiesl2013, KlossnerBecker2013, RobertsRosenthal2024, WallaceHaigh2013}.
%Indeed, as the official description of the draw procedure \citep{FIFA2025b} admits, the 2026 FIFA World Cup draw ``retains the random element'' only \emph{insofar as possible}, see the citation above.
Five measures are considered to quantify this non-uniformity (see Section~\ref{Sec35}). 

Our second main contribution resides in investigating whether the draw procedure of the 2026 FIFA World Cup can be improved with respect to these objective functions. We examine using a different order of the four pots from which the teams are drawn, as well as another group labelling policy, because the non-uniformity of both the Skip \citep{Csato2025c} and the Drop \citep{Guyon2017b} mechanisms depends on the order of the pots, and the pre-assignment of the host(s) can also be a source of non-uniformity \citep{Csato2027, RobertsRosenthal2024}. These adjustments provide 47 reasonable alternative draw procedures, which are compared in terms of their non-uniformity (see Section~\ref{Sec4}).
Even though \citet{Csato2025c} has already suggested considering the $4! = 24$ orders of the four pots in the group draw of the FIFA World Cup, \citet{Csato2025c} does not investigate the role of group labelling, and the results for the 2018 FIFA World Cup do not carry over to the 2026 FIFA World Cup due to the higher number of teams, the different set of geographical constraints, and a new type of draw restrictions.

\section{Related literature} \label{Sec2}

Fairness has several different interpretations in the design of sports rules. For instance, a shooting sequence is fair if identical teams have the same chance of winning a shootout \citep{LambersSpieksma2021}, and a single-elimination tournament is fair if stronger teams are more likely to reach every stage of the tournament \citep{PrinceColeSmithGeunes2013}.

Regarding a group draw, \citet{Guyon2015a} calls a procedure fair if it is not biased against any team, that is, no team has a greater chance to end up in a tough group than its peers.
Both empirical \citep{LapreAmato2025, LaprePalazzolo2022, LaprePalazzolo2023} and theoretical \citep{CeaDuranGuajardoSureSiebertZamorano2020, LalienaLopez2019, LalienaLopez2025} papers characterise the FIFA World Cup draw as fair if the groups are balanced. Indeed, if the groups are perfectly balanced, the procedure cannot be biased against any team (although the converse does not necessarily hold).

Another plausible requirement towards a group draw is \emph{uniformity}: it should have uniform distribution over all valid assignments, that is, all feasible outcomes of the draw need to occur with equal probability.
\citet{Jones1990} analysed the uniformity of the 1990 FIFA World Cup draw. Here, the two seeded South American teams -- Argentina and Brazil -- were automatically assigned to Group B and Group C, respectively. Pot 2 contained two South American teams (Colombia, Uruguay) that were not allowed to play in these groups, as well as four UEFA teams (Czechoslovakia, Ireland, Romania, Sweden). Each of these European teams has a probability of 1/4 (1/4) to be assigned to Group B (C) in a uniform draw. In addition, since there were two South American and two UEFA teams from Pot 2 for Groups A and D--F, containing the four seeded UEFA teams, a seeded UEFA team should have a chance of 1/4 (1/8) to play against a given South American (UEFA) team from Pot 2.

However, FIFA decided to assign the six teams from Pot 2 to Group A, containing the seeded team Italy, with an equal probability of 1/6.
If a South American team from Pot 2 had been placed in the same group as Argentina or Brazil, it would have been automatically reallocated to the next group without a South American team. Hence, West Germany (Pot 1) -- automatically assigned to Group D -- played against UEFA teams from Pot 2 with a cumulative probability of only $4/6 \cdot 3/5 \cdot 2/4 \cdot 1/3 + 2/6 \cdot 4/5 \cdot 3/4 \cdot 2/3 = 1/5$, instead of $1/4 + 1/4 = 1/2$ in a uniform draw.

The 2006 FIFA World Cup draw used an analogous procedure to ensure geographic diversity in the group stage. According to \citet{RathgeberRathgeber2007}, the seeded Germany had a chance of 64.29\% to play against one of the two South American teams (Ecuador, Paraguay) drawn from Pot 2, in contrast to all other seeded UEFA teams such as Italy, which had a chance of only 27.14\%, although these probabilities should be equal in a uniform draw. This was a serious flaw because the two South American teams were considered to be stronger than the other six teams in Pot 2. In Section~\ref{Sec31}, we will see a similar issue regarding Pot 4 in the 2026 FIFA World Cup draw.
%Last but not least, the non-uniform distribution of the 2014 FIFA World Cup draw is shown by \citet{Guyon2015a}.

After \citet{Guyon2015a} identified several problems in the 2014 FIFA World Cup draw, including non-uniform distribution, FIFA decided to reform the draw by allocating the teams into pots based on the FIFA World Ranking, and using the Skip mechanism to enforce draw constraints \citep{Guyon2018d}.
Nonetheless, the rules of the 2018 FIFA World Cup draw still implied that some feasible allocations occur with a higher probability. \citet{Csato2025c} computed this non-uniformity under the 24 possible orders of the four pots via Monte Carlo simulations. Even though the official draw order (Pot 1, Pot 2, Pot 3, Pot 4) was the best according to the sum of absolute differences in assignment probabilities for all team pairs, it changed the probability of qualification by more than one percentage point for two countries. Analogously, \citet{RobertsRosenthal2024} calculated the deviation from a uniform draw for the 2022 FIFA World Cup draw, but the authors did not investigate the role of draw order.

%Naturally, uniformity can also be achieved through other draw procedures.
\citet{Guyon2015a} develops a new tractable procedure for the 2014 FIFA World Cup draw that creates eight random, balanced, and geographically diverse groups, is not biased against any team, and makes all outcomes equally likely.
\citet{RobertsRosenthal2024} present two uniformly distributed draw procedures using balls and bowls to guarantee a nice television show that can be tried at \url{http://probability.ca/fdraw/}. However, in contrast to the method of \citet{Guyon2015a}, they involve computer simulations at some point, which threatens transparency.

The use of integer programming has a long history in sports scheduling \citep{Trick2003, vanDoornmalenHojnyLambersSpieksma2023}. It has also been recently applied to the draw of tournaments.
\citet{CeaDuranGuajardoSureSiebertZamorano2020} propose a two-step mixed-integer linear programming model to create balanced groups for the FIFA World Cup. In the first step, it minimises the difference between the maximum and the minimum (absolute or relative) ranking sums, while ensuring the geographic constraints of the draw. In the second step, the difference between the maximum and minimum range is minimised across all groups.
An integer programming approach has been used to simulate the UEFA Champions League league phase draw by \citet{DevriesereGoossensSpieksma2026} and \citet{GuyonBenSalemBuchholtzerTanre2026}. But this draw employs a different procedure, the Drop mechanism, which has already received serious attention in the case of the UEFA Champions League Round of 16 draw \citep{Kiesl2013, KlossnerBecker2013, WallaceHaigh2013, BoczonWilson2023, Csato2025f}.

\section{Methodology} \label{Sec3}

Section~\ref{Sec31} presents the draw constraints used in the 2026 FIFA World Cup.
Section~\ref{Sec32} introduces the Skip mechanism, the traditional procedure for a group draw in sports tournaments, and illustrates its sensitivity to the group labels.
The challenges that emerge in simulating the 2026 FIFA World Cup draw are discussed in Section~\ref{Sec33}. We propose an integer programming implementation of the Skip mechanism in Section~\ref{Sec34}.
Last but not least, Section~\ref{Sec35} defines our measures of non-uniformity.

\subsection{The 2026 FIFA World Cup draw} \label{Sec31}

The 2026 FIFA World Cup is the first edition of this competition contested by 48 national teams, allocated into 12 groups of four teams each. In order to avoid the creation of highly imbalanced groups, the teams are seeded into four pots based on the FIFA World Ranking of 19 November 2025, and each group contains one team from every pot. Although this policy does not ensure perfect balance, we do not deal with the issue of group balance since it has been the topic of both recent empirical \citet{LaprePalazzolo2023} and theoretical \citep{CeaDuranGuajardoSureSiebertZamorano2020, LalienaLopez2019, LalienaLopez2025} studies.

\begin{table}[t!]
  \centering
  \caption{The seeding for the 2026 FIFA World Cup draw}
  \label{Table1}
    \rowcolors{1}{gray!20}{}
\begin{threeparttable}
    \begin{tabularx}{\textwidth}{llCll} \toprule \hiderowcolors
    Country & Confederation &       & Country & Confederation \\ \midrule
    \multicolumn{2}{c}{\textbf{Pot 1}} &       & \multicolumn{2}{c}{\textbf{Pot 2}} \\ \bottomrule \showrowcolors
    United States (14) & CONCACAF &       & Croatia (10) & UEFA \\
    Mexico (15) & CONCACAF &       & Morocco (11) & CAF \\
    Canada (27) & CONCACAF &       & Colombia (13) & CONMEBOL \\
    Spain (1) & UEFA  &       & Uruguay (16) & CONMEBOL \\
    Argentina (2) & CONMEBOL &       & Switzerland (17) & UEFA \\
    France (3) & UEFA  &       & Japan (18) & AFC \\
    England (4) & UEFA  &       & Senegal (19) & CAF \\
    Brazil (5) & CONMEBOL &       & Iran (20) & AFC \\
    Portugal (6) & UEFA  &       & South Korea (22) & AFC \\
    Netherlands (7) & UEFA  &       & Ecuador (23) & CONMEBOL \\
    Belgium (8) & UEFA  &       & Austria (24) & UEFA \\
    Germany (9) & UEFA  &       & Australia (26) & AFC \\ \toprule \hiderowcolors
    \multicolumn{2}{c}{\textbf{Pot 3}} &       & \multicolumn{2}{c}{\textbf{Pot 4}} \\ \bottomrule \showrowcolors
    Norway (29) & UEFA  &       & Jordan (66) & AFC \\
    Panama (30) & CONCACAF &       & Cape Verde (68) & CAF \\
    Egypt (34) & CAF   &       & Ghana (72) & CAF \\
    Algeria (35) & CAF   &       & Cura{\c c}ao (82) & CONCACAF \\
    Scotland (36) & UEFA  &       & Haiti (84) & CONCACAF \\
    Paraguay (39) & CONMEBOL &       & New Zealand (86) & OFC \\
    Tunisia (40) & CAF   &       & UEFA Path A (12/32/69/71) & UEFA \\
    Ivory Coast (42) & CAF   &       & UEFA Path B (28/31/43/63) & UEFA \\
    Uzbekistan (50) & AFC   &       & UEFA Path C (25/45/47/80) & UEFA \\
    Qatar (51) & AFC   &       & UEFA Path D (21/44/59/65) & UEFA \\
    Saudi Arabia (60) & AFC   &       & IC Path 1 (56/70/149) & Three \\
    South Africa (61) & CAF   &       & IC Path 2 (58/76/123) & Three \\ \toprule
    \end{tabularx}
\begin{tablenotes} \footnotesize
\item
\emph{Notes}:
The numbers in parenthesis indicate the rank of the countries according to the November 2025 FIFA World Ranking, underlying the seeding. In the case of play-off winners, the ranks of all countries involved are given.
\item
The winner of IC Path 1 may come from CAF, CONCACAF, or OFC.
This team is not allowed to be in the same group as any nation from these confederations.
\item
The winner of IC Path 2 may come from AFC, CONCACAF, or CONMEBOL. This team is not allowed to be in the same group as any nation from these confederations.
\end{tablenotes}
\end{threeparttable}

\end{table}

The composition of the four pots is shown in Table~\ref{Table1}, together with the confederations of the teams. The three hosts and the nine strongest teams are placed in Pot 1, the next 12 highest-ranked teams are placed in Pot 2, the next 12 are placed in Pot 3, while the six lowest-ranked teams are placed in Pot 4. Pot 4 also contains the four winners of UEFA play-offs and the two winners of intercontinental play-offs that are scheduled to be played only in March 2026, after the draw on 5 December 2025.

Table~\ref{Table1} uncovers that four teams are likely to be much stronger than the other teams in Pot 4: if the highest-ranked teams win the UEFA play-offs, they would fit into Pot 2 or the top of Pot 3 based on the FIFA World Ranking, which could create some strong groups \citep{Johnson2025}. Indeed, seeding the winners of play-offs in the weakest Pot 4 was suboptimal in the 2022 FIFA World Cup draw with respect to group balance \citep{Csato2023d}.

Each group must consist of exactly one team from each pot. Furthermore, three types of draw constraints are imposed \citep{FIFA2025b}:
\begin{itemize}
\item
\emph{Allocation of hosts}:
Mexico is automatically assigned to Group A, Canada to Group B, and the United States to Group D.

\item
\emph{Balanced distribution of the highest-ranked teams}:
The top four seeds, Spain, Argentina, France, and England, are treated differently from the other teams drawn from Pot 1. Spain and Argentina, as well as France and England, should play in opposite pathways in order to guarantee that, if they win their groups, they can only meet in the final. These four teams should also play in different quarters; if they win their respective groups, they can only meet in the semifinals. \\
The quarters and pathways are determined by the labels of the groups. The first pathway consists of the first quarter $\{ E, F, I \}$ and the second quarter $\{ D, G, H \}$. The second pathway consists of the third quarter $\{ A, C, L \}$ and the fourth quarter $\{ B, J, K \}$.

\item
\emph{Geographic separation}:
No group could have more than one team from the same confederation, except for UEFA, which has 16 teams in the 2026 FIFA World Cup. Each group needs to contain at least one, but no more than two UEFA teams. \\
The placeholders of the play-offs are considered for each confederation from which the winner may come.
\end{itemize}
Appendix~\ref{Appendix_A} clarifies the interpretation of the second set of constraints, the distribution of the highest-ranked teams. 

\subsection{The Skip mechanism} \label{Sec32}

Tournament organisers traditionally use the so-called Skip mechanism to create a valid group assignment \citep{Csato2025f}. This works as follows:
\begin{itemize}
\item
The order of the pots from which the teams are drawn is chosen;
\item
The groups are labelled;
\item
The team drawn currently is assigned to the first available group in alphabetic order such that at least one feasible assignment remains for the teams still to be drawn;
\item
The above procedure is repeated until all pots are sequentially emptied in the given order.
\end{itemize}
A simulator of the Skip mechanism for the 2018 and 2022 FIFA World Cup draws is available at \url{http://probability.ca/fdraw/}.
The video of the 2026 FIFA World Cup draw can be found at \url{https://www.youtube.com/watch?v=9HX_tQBA-Iw}.

Consider an example illustrating when a group is skipped.

\begin{example} \label{Examp1}
Belgium is drawn first from Pot 1, and assigned to Group C since Group A is occupied by Mexico and Group B by Canada.
Argentina is drawn second from Pot 1, and assigned to Group E since Group D is occupied by the United States.
Then Spain is drawn third from Pot 1, and assigned to Group J: even though Groups F, G, H, I still have an empty slot for a team from Pot 1, Spain cannot play in the same pathway as Argentina.
\end{example}

Naturally, the problem is not always as simple to solve as in Example~\ref{Examp1}, since the draw needs to avoid any deadlock, when it is no longer possible to complete the draw.

\begin{example} \label{Examp2}
Continue Example~\ref{Examp1}.
Portugal is drawn fourth, and assigned to Group F.
Brazil is drawn fifth, and assigned to Group G.
Germany is drawn sixth, and assigned to Group I -- even though Group H also has an empty slot.
Why? The remaining teams in Pot 1 are France, England, and the Netherlands. Slots are available in Groups H, I, L, and K. France and England cannot be assigned to Group I since they are not allowed to play in the same quarter with Argentina (which plays in Group E). Analogously, they cannot be assigned to Group K because Spain is in Group J. Consequently, the team drawn first (second) from the set of France and England should be in Group H (L); thus, Germany cannot be placed in Group H. 
\end{example}

The assignment probabilities of the Skip mechanism depend on the order of the pots \citep{Csato2025c}.
Even though the increasing order Pot 1, Pot 2, Pot 3, Pot 4 is followed in the 2026 FIFA World Cup draw \citep{FIFA2025b}, any permutation of the four pots can be implemented with minimal modifications to the draw procedure. For instance, a reversed order was chosen for the 2020/21 \citep{UEFA2020d} and the 2022/23 \citep{UEFA2021i} UEFA Nations League league phase draws, while the quite arbitrary order Pot 4, Pot 3, Pot 1, Pot 2 was used in the 2025 World Men's Handball Championship \citep{IHF2020}.

The following example shows that the Skip mechanism is also sensitive to the labels of the groups.

\begin{example} \label{Examp3}
Consider a draw involving two groups where Pot 1 contains teams 1--3 and Pot 2 contains teams 4--6. There is one restriction: teams 2 and 5 cannot be assigned to the same group.

In the absence of the draw constraint, $3! \cdot 3! = 36$ feasible assignments exist if the group labels are taken into account. However, teams 2 and 5 are in the same group in $3 \cdot 2! \cdot 2! = 12$ assignments, which is prohibited. Among the 24 valid assignments, team 1 is placed in the same group as teams 4 and 6 in six cases each (one case each if group labels are ignored), and in the same group as team 5 in 12 cases (two cases if group labels are ignored), since no constraint applies to the remaining four teams 2, 3, 4, and 6.
%Among the remaining 24 assignments, team 1 is in the same group in 6-6 instances with teams 4 and 6 (1-1 if the group labels are ignored), respectively, while in 12 instances with team 5 (2 if the group labels are ignored since no constraint applies for the remaining teams 2, 3, 4, 6). 
Hence, team 1 plays against team 5 with a probability of 1/2, and against teams 4 and 6 each with a probability of 1/4 in a uniform draw.

If team 1 is pre-assigned to Group A, then it has a probability of 1/3 to play against each of teams 4, 5, 6 in the group stage under the Skip mechanism.

Now see the case when team 1 is drawn from Pot 1 randomly.
It is drawn first and placed in Group A with a probability of 1/3.
It is drawn second and placed in Group B with a probability of 1/3. If team 2 is in Group A, which has an overall probability of 1/6, team 1 is assigned to the same group as team 5 with a probability of 2/3 (if team 5 is drawn first or second from Pot 2). Analogously, if team 2 is in Group C, team 1 is assigned to the same group as team 5 with a probability of 2/3 (if team 5 is drawn second or third from Pot 2).
Finally, if team 1 is drawn third and placed in Group C, team 2 is in Group A (B) with an overall probability of 1/6 (1/6), and team 1 is assigned to the same group as team 5 with a probability of 1/3 (2/3).
Therefore, teams 1 and 5 are assigned to the same group with a chance of $1/3 \cdot 1/3 + 1/6 \cdot \left( 2/3 + 2/3 \right) + 1/6 \cdot \left( 1/3 + 2/3 \right) = 1/9 + 2/9 + 1/6 = 1/2$.

To summarise, the Skip mechanism is non-uniform if team 1 is pre-assigned to Group A, but uniform if team 1 is drawn randomly from Pot 1.
\end{example}

According to Example~\ref{Examp3}, pre-assigning a team to a group -- or, more generally, imposing constraints that depend on the group labels -- can influence the non-uniformity of the draw. However, the groups could be labelled accordingly even \emph{after} the group draw. For instance, if the organiser wants to guarantee the assignment of Mexico to Group A, then the group of Mexico is simply called Group A once the group composition is obtained.

Ex-post group labelling can be implemented as follows.
In the first step, the groups are drawn without any constraint on their labels (i.e.\ the constraints on the allocation of hosts and balanced distribution of the highest-ranked teams, presented in Section~\ref{Sec31}, are ignored). Naturally, this requires the assignment of group labels in the second step. The group of Mexico is called A, the group of Canada is called B, and the group of the United States is called D. Balanced distribution of the highest-ranked teams is guaranteed by pairing the remaining nine group labels and the drawn groups in another constrained draw using the Skip mechanism.

While this requires two draws instead of one, the constraints are also separated into two sets, which might be easier for the audience to understand. It is also possible that the second draw, the assignment of group labels, would not be interesting for the public once the group compositions are already known. Then the second draw can be conducted under the supervision of only national team representatives. In a sense, the Union of European Football Associations (UEFA) follows this idea in the draw of its club competitions, where the set of matches to be played in the incomplete round-robin league is drawn publicly, but the exact schedule (order) of the matches is determined later, behind closed doors \citep{GuyonBenSalemBuchholtzerTanre2026}.

Based on these arguments, the official draw procedure for the 2026 FIFA World Cup has 47 reasonable alternatives by changing the order of the pots and labelling the groups ex-ante (before the draw) or ex-post (after the draw).
They will be compared and evaluated in Section~\ref{Sec4} with respect to their non-uniformity, quantified by the five measures that will be introduced in Section~\ref{Sec35}.

\subsection{Challenges of simulating the 2026 FIFA World Cup draw} \label{Sec33}

The Skip mechanism should consider $9! \cdot \left( 12! \right)^3 \approx 3.99 \cdot 10^{31}$ scenarios for each order of the pots with ex-ante group labelling, and $\left( 12! \right)^4 \approx 5.26 \cdot 10^{34}$ scenarios for each order of the pots with ex-post group labelling.
Due to this enormous number of scenarios, the enumeration of all possible draws is computationally impractical. We therefore use Monte Carlo simulation as a tractable approach to generate random draws according to the official Skip mechanism.

Based on our experience with the 2018 \citep{Csato2025c} and 2022 \citep{Csato2023d} FIFA World Cup draws, this computation was expected to be trivial. Indeed, both \citet{RobertsRosenthal2024} and \citet{Csato2025c} suggested recursive backtracking algorithms to simulate the Skip mechanism, which produces one million random draws for any order of the pots in about 1-2 hours for these examples.

\begin{definition} \label{Def1}
\emph{Recursive backtracking algorithm}:
Given a partial draw and a random order in which the teams are drawn from the pots, it assigns the next team to the next available slot and checks \emph{all} permutations of the remaining teams to see whether the draw can be completed. If yes, the randomly selected team is assigned accordingly. If not, it attempts to assign this team to the subsequent available slot, and so on.
\end{definition}

\begin{figure}[t!]
\centering

\begin{tikzpicture}[scale=1,auto=center, transform shape, >=triangle 45]
\tikzstyle{every node}=[draw,align=center];
  \node (N0) at (-2,14) {Initialisation \\ $i = 1$};
  \node (N1) at (-2,12) {START \\ $j = 1$};
  \node[shape = ellipse] (N2) at (-2,9) {CHECK \\ Can team $i$ be assigned \\ to group $j$?};
  \node (N3) at (-2,6) {Team $i$ is \\ assigned to group $j$};
  \node[shape = ellipse] (N4) at (-2,3) {Is $i$ smaller than the \\ number of teams?};
  \node (N5) at (-5,0) {$i = i + 1$ \\ Go to START};
  \node (N6) at (-1.5,-0.78) {\textbf{\textcolor{ForestGreen}{The feasible}} \\ \textbf{\textcolor{ForestGreen}{allocation implied}} \\ \textbf{\textcolor{ForestGreen}{by the Skip}} \\ \textbf{\textcolor{ForestGreen}{mechanism is}} \\ \textbf{\textcolor{ForestGreen}{found}}};
  \node[shape = ellipse] (N7) at (6,6) {Is $j$ smaller than the \\ number of groups?};
  \node (N8) at (3,3) {$j = j + 1$ \\ Go to CHECK};
  \node[shape = ellipse] (N9) at (7,3) {Is $i = 1$?};
  \node (N10) at (2.25,-0.28) {\textbf{\textcolor{BrickRed}{There exists}} \\ \textbf{\textcolor{BrickRed}{no feasible}} \\ \textbf{\textcolor{BrickRed}{allocation}}};
  \node (N11) at (7,-0.78) {$j = \left[ \text{group of team } (i-1) \right] + 1$ \\ All teams $k \geq i$ are \\ removed from their groups \\ $i = i - 1$ \\ Go to START};

\tikzstyle{every node}=[align=center];  
  \draw [->,line width=1pt] (N0) -- (N1)  node [midway, left] {};
  \draw [->,line width=1pt] (N1) -- (N2)  node [midway, left] {};
  \draw [->,line width=1pt] (N2) -- (N3)  node [midway, left] {Yes};
  \draw [->,line width=1pt] (N3) -- (N4)  node [midway, left] {};
  \draw [->,line width=1pt] (N4) -- (N5)  node [midway, above left] {Yes};
  \draw [->,line width=1pt] (N4) -- (N6)  node [midway, above right] {No};
  \draw [->,line width=1pt] (N2) -- (N7)  node [midway, above right] {No};
  \draw [->,line width=1pt] (N7) -- (N8)  node [midway, above left] {Yes};
  \draw [->,line width=1pt] (N7) -- (N9)  node [midway, above right] {No};
  \draw [->,line width=1pt] (N9) -- (N10)  node [midway, above left] {Yes};
  \draw [->,line width=1pt] (N9) -- (N11)  node [midway, right] {No};
%  \draw [->,line width=1pt] (N11) -- (N12)  node [midway, left] {};
\end{tikzpicture}
\captionsetup{justification=centering}
\caption{A recursive backtracking algorithm that finds the feasible \\ group composition corresponding to a given random order of the teams}
\label{Fig1}
\end{figure}

%\end{document}

Figure~\ref{Fig1} shows a recursive backtracking algorithm. It can reliably detect any subsequent conflict, no matter where the conflict occurs -- but requires trying all possible permutations of the teams still to be drawn in case of a conflict.

A recursive backtracking algorithm is attractive because it can be easily implemented in any program, without any other tool such as a linear programming solver. However, it struggles with the 2026 FIFA World Cup draw. 

This can be explained by the quite restrictive constraints of the 2026 FIFA World Cup draw. In Pot 4, the placeholders of the UEFA play-offs are treated analogously to UEFA teams in the other pots. The winner of the first inter-continental play-off cannot be assigned to a group with a CAF, CONCACAF, or OFC team. The first three pots contain seven CAF and four CONCACAF teams; hence, at least one group remains available for this placeholder even if all constraints related to IC Path 1 are ignored until Pot 3 is emptied.

However, the winner of the second inter-continental play-off can come from AFC, CONCACAF, or CONMEBOL, which gives $5 + 7 + 5 = 17$ deadlocking nations in Pots 1--3, while there are 12 groups. Thus, the draw can be completed only if the placeholder of IC Path 2 is assigned to a group with one CAF and two UEFA countries (see \url{https://x.com/julienguyon1977/status/1994785539523056034?s=20} and \citet{Johnson2025}).
Interestingly, this constraint became effective in the 2026 FIFA World Cup draw: Uzbekistan was not assigned to the first empty slot in Group I, which contained France and Senegal at that moment, in order to guarantee a place for the winner of IC Path 2 (\url{https://x.com/julienguyon1977/status/1997018050067833231?s=20}). Consequently, its group was fixed even before the draw reached Pot 4.

\begin{claim} \label{Claim1}
Simulating the 2026 FIFA World Cup is intractable with a recursive backtracking algorithm.
\end{claim}

\begin{table}[t!]
  \centering
  \caption{A possible start of the 2026 FIFA World Cup draw if \\ the constraints for the teams drawn from Pots 3--4 are ignored}
  \label{Table2}
    \rowcolors{1}{gray!20}{}
    \begin{tabularx}{0.9\textwidth}{l LLLL} \toprule \hiderowcolors
    \multirow{2}[0]{*}{Group} & \multicolumn{2}{c}{Pot 1} & \multicolumn{2}{c}{Pot 2} \\
          & Country & Confederation & Country & Confederation \\ \bottomrule \showrowcolors
    A     & Mexico & CONCACAF & Morocco & CAF \\
    B     & Canada & CONCACAF & Austria & UEFA \\
    C     & Belgium & UEFA  & Australia & AFC \\
    D     & United States & CONCACAF & Croatia & UEFA \\
    E     & England & UEFA  & Iran  & AFC \\
    F     & France & UEFA  & Japan & AFC \\
    G     & Argentina & CONMEBOL & Switzerland & UEFA \\
    H     & Brazil & CONMEBOL & Senegal & CAF \\
    I     & Germany & UEFA  & South Korea & AFC \\
    J     & Netherlands & UEFA  & Colombia & CONMEBOL \\
    K     & Portugal & UEFA  & Ecuador & CONMEBOL \\
    L     & Spain & UEFA  & Uruguay & CONMEBOL \\ \toprule
    \end{tabularx}
\end{table}

%\begin{proposition} \label{Prop1}
%A recursive backtracking algorithm is highly unlikely to generate a sufficient number of %random draws for the 2026 FIFA World Cup.

\begin{proof}
We show that a recursive backtracking algorithm is extremely unlikely to generate a sufficient number of random draws for the 2026 FIFA World Cup.

Consider the situation presented in Table~\ref{Table2}, a possible assignment of the teams from Pots 1--2 to the 12 groups by ignoring all restrictions that apply to the teams drawn from Pots 3--4. This draw cannot be completed since no group contains either two UEFA teams, or one CAF and one UEFA team. %Therefore, the Skip mechanism could not assign Senegal to Group H, but only to Group I. 
A recursive backtracking algorithm is able to recognise the problem only after checking all possible orders of the 24 teams drawn from Pots 3--4. This requires $\left( 12! \right)^{2} \approx 2.29 \cdot 10^{17}$ steps, which is roughly the total number of draws in the 2018 and 2022 FIFA World Cups. Hence, recursive backtracking is guaranteed to get stuck, even on a supercomputer, if it encounters a situation analogous to Table~\ref{Table2}.

By simulating random draws from Pots 1--2, we have estimated the probability of an analogous scenario that turns out to be about $p = 0.2315\%$. Therefore, a recursive backtracking algorithm can avoid the problem of Table~\ref{Table2} with a chance of merely $(1-p)^{10,000} \approx 8.6 \cdot 10^{-11}$ if the draw is simulated only 10 thousand times, but this is insufficient for a robust approximation of the assignment probabilities.
\end{proof}

This barrier can perhaps be solved by devising appropriate constraints that can recognise a deadlock analogous to Table~\ref{Table2} earlier. However, such a manual modification would be burdensome and depend on the set of draw constraints. Consequently, simulating the 2026 FIFA World Cup draw requires a different approach.

\subsection{A novel implementation using integer programming} \label{Sec34}

%Here we discuss how the draw can be simulated by using integer programming.

Let $T$ be the set of teams, $G$ be the set of groups, $P$ be the set of pots, and $C$ be the set of confederations. Let $T_c$ be the set of teams from confederation $c \in C$, and let $T_p$ be the set of teams drawn from Pot $p \in P$.
We define the following sets based on Section~\ref{Sec31}:
\begin{itemize}
\item
$Q_1 = \{E,F,I\}, \;
Q_2 = \{D,G,H\}, \;
Q_3 = \{A,C,L\}, \;
Q_4 = \{B,J,K\}$;

\item
$H_1 = Q_1 \cup Q_2, \quad
H_2 = Q_3 \cup Q_4$;

\item
$T^\star = \{\text{Spain},\text{Argentina},\text{France},\text{England}\}$;

\item
$\mathcal{F} = \big\{ \{\text{Spain},\text{Argentina}\},\{\text{France},\text{England}\} \big\}$.
\end{itemize}
Set $Q_k$ represents the $k$th quarter of the knockout bracket. Set $H_1$ ($H_2$) is the first (second) pathway, consisting of the first and second (third and fourth) quarters. The four highest-ranked national teams are given by set $T^\star$.

Any draw is determined by $48 \cdot 12 = 576$ binary variables $x_{i,g}$, which equals 1 if team $i \in T$ is assigned to group $g \in G$, and 0 otherwise.
A feasible draw should satisfy the following integer program:

\begin{align}
    & \sum_{i \in T_{c}}x_{i,g} \leq 1 & \forall g \in G, \ \forall c \in C \setminus \{ \text{UEFA} \} \label{c1} \\
    & \sum_{i \in T_{ \text{UEFA} }}x_{i,g} \geq 1 & \forall g \in G \label{c2}\\
    & \sum_{i \in T_{ \text{UEFA} }}x_{i,g} \leq 2 & \forall g \in G \label{c3}\\
    & \sum_{g \in G}x_{i,g} = 1 & \forall i \in T \label{c4} \\
    & \sum_{i \in T_p}x_{i,g} = 1 & \forall g \in G, \ \forall p \in P \label{c5} \\
    & \sum_{g \in H_k} x_{i,g} + \sum_{g \in H_k} x_{j,g} \leq 1 
    & \forall \{i,j\} \in \mathcal{F}, \ \forall k \in \{1,2\} \label{c6}\\
    & \sum_{g \in Q_k} \sum_{i \in T^\star} x_{i,g} \leq 1
    & \forall k \in \{1,2,3,4\} \label{c7} \\
    & x_{\text{Mexico},\text{A}} = x_{\text{Canada},\text{B}} = x_{\text{United States},\text{D}} = 1 \label{c8} \\
    & x_{i,g} \in \{0,1\} & \forall i \in T, \ \forall g \in G \label{c9}
\end{align}
Constraints~\eqref{c1} state that no group contains more than one team from the same confederation, except for UEFA. Constraints~\eqref{c2}--\eqref{c3} enforce that every group contains at least one and at most two teams from UEFA. Constraints~\eqref{c4} ensure that every team is assigned to exactly one group, and constraints~\eqref{c5} guarantee that each group contains exactly one team from each pot. Constraints~\eqref{c6} imply that, for every pair in $\mathcal{F}$, the two teams should be in opposite pathways. Constraints~\eqref{c7} enforce that the four teams in $T^{\star}$ play in different quarters. Constraints~\eqref{c8} show the pre-assignment of the three host nations. Finally, expressions~\eqref{c9} give the domains of the variables.

If the official ex-ante group labelling is used, we directly follow the Skip mechanism described in Section~\ref{Sec32}.
An unallocated team is drawn randomly from the actual pot. It is checked whether assigning this team to the first available group, in alphabetical order, allows a solution to constraints \eqref{c1}--\eqref{c9}. If yes, the given team is assigned to this group (the corresponding variable is set to 1), and we draw a new team from the set of unassigned teams. If no feasible solution exists by assigning the given team to this group, the draw cannot be completed without violating a constraint now or later. Then, it is checked whether assigning the given team to the next available group results in a feasible solution.
Analogously to the Skip mechanism, this procedure is repeated until the given team can be assigned to a group such that the draw can be completed -- which is possible, since the assignment of the previous team was verified to leave at least one feasible outcome. Hence, the corresponding binary variable is fixed at 1, and the next unallocated team is drawn according to the fixed order of the pots, until all teams are assigned to a group.

If the group labelling is done only ex-post, constraints~\eqref{c6} and \eqref{c7} are ignored because they can be satisfied by appropriately labelling the groups containing the four teams in set $T^{\star}$. Similarly, there is no need for constraints~\eqref{c8}, and the three hosts are treated similarly to the other nine teams drawn from Pot 1.

\begin{figure}[t!]
\centering

\begin{tikzpicture}[scale=1,auto=center, transform shape, >=triangle 45]
\tikzstyle{every node}=[draw,align=center];
  \node (N0) at (-2,17) {IP Setup \\ $i=1$};
  \node (N1) at (-2,15) {START \\ $j=1$};
  %\node[shape = ellipse] (N2) at (-2,9) {Can team $i$ of pot $j$ be \\ assigned to group $k$?};
  \node (N10) at (-2,12) {UPDATE \\ Add a new constraint \\ to the IP: Team $i$ is \\ assigned to group $j$};
  \node[shape = ellipse] (N2) at (-2,9) {Does the IP have a \\ feasible solution?};
  \node (N3) at (-2,6) {Team $i$ is \\ assigned to group $j$};
  \node[shape = ellipse] (N4) at (-2,3) {Is $i$ smaller than the \\ number of teams?};
  \node (N5) at (-5,0) {$i = i + 1$ \\ Go to START};
  \node (N6) at (-1.5,-0.78) {\textbf{\textcolor{ForestGreen}{The feasible}} \\ \textbf{\textcolor{ForestGreen}{allocation implied}} \\ \textbf{\textcolor{ForestGreen}{by the Skip}} \\ \textbf{\textcolor{ForestGreen}{mechanism is}} \\ \textbf{\textcolor{ForestGreen}{found}}};
  \node[shape = ellipse] (N7) at (6,6) {Is $j$ smaller than the \\ number of groups?};
  \node (N8) at (4,2.5) {Remove the constraint \\ that team $i$ is assigned \\ to group $j$ from the IP};
  \node (N11) at (4,-1) {$j = j + 1$ \\ Go to UPDATE};
  \node (N9) at (8,2.5) {\textbf{\textcolor{BrickRed}{There exists}} \\ \textbf{\textcolor{BrickRed}{no feasible}} \\ \textbf{\textcolor{BrickRed}{allocation}}};

\tikzstyle{every node}=[align=center]; 
  \draw [->,line width=1pt] (N0) -- (N1)  node [midway, left] {}; 
  \draw [->,line width=1pt] (N1) -- (N10)  node [midway, left] {};
  \draw [->,line width=1pt] (N10) -- (N2)  node [midway, left] {};
  \draw [->,line width=1pt] (N2) -- (N3)  node [midway, left] {Yes};
  \draw [->,line width=1pt] (N3) -- (N4)  node [midway, left] {};
  \draw [->,line width=1pt] (N4) -- (N5)  node [midway, above left] {Yes};
  \draw [->,line width=1pt] (N4) -- (N6)  node [midway, above right] {No};
  \draw [->,line width=1pt] (N2) -- (N7)  node [midway, above right] {No};
  \draw [->,line width=1pt] (N7) -- (N8)  node [midway, above left] {Yes};
  \draw [->,line width=1pt] (N7) -- (N9)  node [midway, above right] {No};
  \draw [->,line width=1pt] (N8) -- (N11)  node [midway, above right] {};
%  \draw [->,line width=1pt] (N9) -- (N10)  node [midway, above left] {Yes};
%  \draw [->,line width=1pt] (N9) -- (N11)  node [midway, right] {No};
%  \draw [->,line width=1pt] (N11) -- (N12)  node [midway, left] {};
\end{tikzpicture}
\captionsetup{justification=centering}
\caption{An integer programming-based algorithm that finds the feasible \\ group composition corresponding to a given random order of the teams}
\label{Fig2}
\end{figure}

%\end{document}

Figure~\ref{Fig2} shows the corresponding algorithm based on integer programming. There are two main differences compared to the recursive backtracking algorithm in Figure~\ref{Fig1}:
(1) the integer program above checks whether team $i$ can be assigned to group $j$; consequently,
(2) if the answer is negative, either there is still a group with an empty slot for team $i$, or no feasible solution exists.
To summarise, the backtracking part can be removed as the integer program immediately detects any conflict for the teams still to be drawn. Naturally, this simplification has the price of solving the integer program first. However, according to Claim~\ref{Claim1}, this is necessary to avoid the occurrence of an excessively long backtracking loop with a non-marginal probability.

1.2 million random draws were simulated for each of the 48 possible variants of the Skip mechanism considered, given by the 24 orders of the pots and the two group labelling policies. We used a GNU/Linux-based system with an AMD EPYC 7532 32-Core Processor running at 3.3GHz, provided with 8 threads and 64GB of RAM. The integer programming models were solved by Gurobi 12.0 (\url{https://www.gurobi.com/}).
This resulted in a computation time of about 14 hours on average for each of the 48 draw procedures, corresponding to less than 0.05 seconds on average per draw. Nonetheless, the worst-case could be more relevant: the maximum runtime was 3.09 (1.30) seconds for the $24 \cdot 1.2 = 28.8$ million random draws with ex-ante (ex-post) group labelling. Consequently, the proposed method can be implemented in a live setting, during the actual TV broadcast.

\subsection{Quantifying the non-uniformity of a draw procedure} \label{Sec35}

A uniform draw selects any feasible allocation of the teams into groups with equal probability, which can be achieved by a \emph{rejection sampler} \citep{Guyon2014a, Guyon2015a, RobertsRosenthal2024}. In the first step, a draw is generated uniformly and randomly without the draw constraints. In the second step, this assignment is accepted if all restrictions are satisfied, and rejected if at least one restriction is violated.
However, checking all possible cases in the 2026 FIFA World Cup is hopeless, as the number of draws equals $9! \cdot \left( 12! \right)^3 \approx 3.99 \cdot 10^{31}$.
%Therefore, the rejection sampler is used repeatedly until the required number of draws is obtained.

In the case of the 2018 FIFA World Cup, about one out of 161 random draws is accepted by the rejection sampler \citep{Csato2025c}. The analogous ratio for the 2022 FIFA World Cup draw is one out of 560 \citep{RobertsRosenthal2024}.
However, for the 2026 FIFA World Cup, the rejection sampler accepts only about one out of 1 million random draws due to the highly restrictive set of draw constraints. We generated 100 million uniformly distributed valid assignments, using R \citep{R2025} with the additional packages \texttt{doRNG} \citep{Gaujoux2025} and \texttt{kableExtra} \citep{Zhu2024}. This took just under 11 days on 100 parallel threads.

Given the size of the outcome space, it is not feasible to observe each possible outcome even once in simulation, let alone sufficiently often to estimate its probability of occurrence. Therefore, the analysis must rely on suitable functionals of the distribution, such as the probability that two teams $i$ and $j$ are assigned to the same group.

Let $p_{ij}^U$ ($p_{ij}^D$) be the probability of assigning teams $i$ and $j$ to the same group under a uniform draw $U$ (a given draw procedure $D$). The differences between $p_{ij}^U$ and $p_{ij}^D$ will be aggregated to measure the non-uniformity of $D$.
$i < j$ is assumed in the following to avoid double counting.

If seeding constraints are imposed without any additional restriction, each team has 36 possible opponents in the group stage, which leads to 864 team pairs with a positive probability of occurring. If a confederation (except for UEFA) has $p$ and $q$ teams in two pots, respectively, $p \cdot q$ team pairs are prohibited. Therefore, the number of team pairs that cannot play against each other in the group stage is $4 \cdot 3 + 4 \cdot 2 + 3 \cdot 2 = 26$ for AFC, $2 \cdot 5 + 2 \cdot 3 + 5 \cdot 3 = 31$ for CAF, $3 \cdot 1 + 3 \cdot 4 + 1 \cdot 4 = 19$ for CONCACAF, and $2 \cdot 3 + 2 \cdot 1 + 2 \cdot 1 + 3 \cdot 1 + 3 \cdot 1 + 1 \cdot 1 = 17$ for CONMEBOL. It is zero for OFC since the only OFC team and the winner of the first intercontinental play-off are both assigned to Pot 4. Finally, any two UEFA teams can be assigned to the same group if they are drawn from different pots.

Thus, the number of team pairs with a non-zero probability, $P_{>0} = \# \{ p_ {ij}: p_ {ij}^U > 0 \}$, equals $864 - 26 - 31 - 19 - 17 = 771$. Indeed, the $N^U = 100{,}000{,}000$ uniform draws, as well as the $N^D= 1{,}200{,}000$ draws for any given draw procedure, reinforce that 771 positive pairing probabilities exist; hence, a combination of constraints does not prevent additional pairings. Under a uniform draw, the minimum probability that two given teams are assigned to the same group is 2.01\% (for the UEFA teams in Pots 2 and 3, e.g.\ Croatia and Norway), while the maximum probability reaches 24.43\% (for the UEFA teams in Pot 2 and the winner of IC Path 2).

Denote the set of all teams by $T = \{ i: 1 \leq i \leq 48 \}$, the set of teams in Pot 1 by $T_1 = \{ i: 1 \leq i \leq 12 \}$, the set of teams in Pots 1--3 by $T_{123} = \{ i: 1 \leq i \leq 36 \}$, and the set of UEFA teams drawn from Pot 4 by $T_{4U} = \{ i: 43 \leq i \leq 46 \}$, respectively.
Denote the $k$th highest element of a set by superscript $(k)$.

We consider five different metrics to quantify the non-uniformity of a draw procedure:
\begin{itemize}
\item
The mean of absolute differences in assignment probabilities for all team pairs with a non-zero probability:
\[
M_1 = \frac{\sum \lvert p_{ij}^D - p_{ij}^U \rvert: i,j \in T}{P_{>0}};
\]
\item
The highest absolute difference in assignment probabilities for all team pairs:
\[
M_2 = \left\{ \lvert p_{ij}^D - p_{ij}^U \rvert: i,j \in T \right\}^{(1)};
\]
\item
The mean of the eight highest differences in assignment probabilities for all team pairs:
\[
M_3 = \frac{\sum_{k=1}^8 \left\{ \lvert p_{ij}^D - p_{ij}^U \rvert: i,j \in T \right\}^{(k)}}{8};
\]
\item 
The mean of absolute differences in assignment probabilities for all team pairs containing a UEFA team drawn from Pot 4:
\[
M_4 = \frac{\sum \lvert p_{ij}^D - p_{ij}^U \rvert: i \in T_{123} \text{ and } j \in T_{4U}}{36 \cdot 4};
\]
\item
The mean of absolute differences in assignment probabilities for all team pairs containing a team drawn from Pot 1 and a UEFA team drawn from Pot 4:
\[
M_5 = \frac{\sum \lvert p_{ij}^D - p_{ij}^U \rvert: i \in T_{1} \text{ and } j \in T_{4U}}{12 \cdot 4}.
\]
\end{itemize}
All measures $M_1$--$M_5$ are multiplied by 100 to get them in percentage points. Crucially, none of these metrics use exogenous information such as team rankings or Elo ratings; they only depend on the characteristics of the draw procedure $D$.

$M_1$ is perhaps the most convenient way of aggregation, but the average may mask large changes for individual teams. $M_2$ focuses on the largest deviation from a uniform draw, while $M_3$ considers the eight largest differences in assignment probabilities.
Finally, $M_4$ and $M_5$ are motivated by the strange policy of assigning the four winners of UEFA play-offs to the weakest Pot 4. As a result, each team in Pots 1--3 has a strong preference to avoid a UEFA team from Pot 4. Any difference in these assignment probabilities may be substantially more costly for the organiser than one affecting other team pairs, which motivates the use of $M_4$. Analogously, $M_5$ restricts attention to the teams in Pot 1 and Pot 4, by assuming that the strongest teams and the hosts should primarily be ``defended'' against the imbalance existing in Pot 4. In the formulas of $M_4$ and $M_5$, no difference in assignment probabilities equals zero by definition, since every team in Pots 1--3 can play against a UEFA team in Pot 4 with a non-zero probability.

For a given pair of teams $i$ and $j$, the standard error of $\lvert p_{ij}^D - p_{ij}^U \rvert$ equals:
\[
\sqrt{\frac{p_{ij}^D \left( 1 - p_{ij}^D \right)}{N^D = 1{,}200{,}000} + \frac{p_{ij}^U \left( 1 - p_{ij}^U \right)}{N^U = 100{,}000{,}000}}.
\]
Since both $p_{ij}^D \left( 1 - p_{ij}^D \right)$ and $p_{ij}^U \left( 1 - p_{ij}^U \right)$ have a conservative upper bound of 0.25, the standard error is at most 0.046\%. Hence, the 99\% confidence interval for $\lvert p_{ij}^D - p_{ij}^U \rvert$ is given by $\lvert p_{ij}^D - p_{ij}^U \rvert \pm 0.118\%$.

Note also that teams from the same confederation and seeding pot are interchangeable and have the same probability of playing against any opponent, except for Argentina, England, France, and Spain due to the competitive balance constraint applying to these teams. However, England and France are also interchangeable.

%Finally, note that neither $p_{ij}^D$ under the draw procedure studied nor the corresponding benchmark value $p_{ij}^U$ under the uniform distribution could be computed exactly, since they depend on the set of feasible outcomes, which is itself unknown. Therefore, both sets of probabilities are estimated by Monte Carlo simulations as discussed in Section~\ref{Sec34} and \ref{Sec35}.

\section{Results} \label{Sec4}

\begin{table}[t!]
\centering
\caption{Measures of non-uniformity in the 2026 FIFA World Cup draw \\ under all draw orders of the Skip mechanism: ex-ante group labelling}
\label{Table3}

\centerline{
\begin{threeparttable}
\rowcolors{3}{}{gray!20}
    \begin{tabularx}{1.2\linewidth}{l Rrr Rrr Rrr Rrr Rrr} \toprule \hiderowcolors
    Measure & \multicolumn{3}{c}{Average ($M_1$)} & \multicolumn{3}{c}{Maximum ($M_2$)} & \multicolumn{3}{c}{Maximum 8 ($M_3$)} & \multicolumn{3}{c}{Pot 4 ($M_4$)} & \multicolumn{3}{c}{Pots 1/4 ($M_5$)} \\
    Order & Value  & \multicolumn{2}{c}{Chg.} & Value  & \multicolumn{2}{c}{Chg.} & Value  & \multicolumn{2}{c}{Chg.} & Value  & \multicolumn{2}{c}{Chg.} & Value  & \multicolumn{2}{c}{Chg.} \\ \bottomrule \showrowcolors
    1-2-3-4 & 0.64  & \multicolumn{2}{c}{---} & 3.85  & \multicolumn{2}{c}{---} & 3.51  & \multicolumn{2}{c}{---} & 0.70  & \multicolumn{2}{c}{---} & 1.64  & \multicolumn{2}{c}{---} \\
    1-2-4-3 & 0.97  & \up & 50\%  & 15.83 & \up & 311\% & 8.70  & \up & 148\% & 0.97  & \up & 311\% & 1.36  & \down & 17\% \\
    1-3-2-4 & 0.71  & \up & 10\%  & 4.52  & \up & 17\%  & 4.05  & \up & 15\%  & 0.73  & \up & 17\%  & 1.68  & \up & 2\% \\
    1-3-4-2 & 1.09  & \up & 69\%  & 8.70  & \up & 126\% & 6.62  & \up & 89\%  & 0.98  & \up & 126\% & 1.64  & \up & 0\% \\
    1-4-2-3 & 1.18  & \up & 83\%  & 16.52 & \up & 329\% & 9.06  & \up & 158\% & 1.59  & \up & 329\% & 3.19  & \up & 94\% \\
    1-4-3-2 & 1.23  & \up & 91\%  & 6.40  & \up & 66\%  & 6.21  & \up & 77\%  & 1.59  & \up & 66\%  & 3.19  & \up & 94\% \\ \hline
    2-1-3-4 & 0.64  & \up & 0\%   & 3.92  & \up & 2\%   & 3.52  & \up & 0\%   & 0.72  & \up & 2\%   & 1.62  & \down & 1\% \\
    2-1-4-3 & 0.95  & \up & 48\%  & 15.38 & \up & 299\% & 8.79  & \up & 150\% & 0.99  & \up & 299\% & 1.39  & \down & 15\% \\
    2-3-1-4 & 0.71  & \up & 11\%  & 3.92  & \up & 2\%   & 3.42  & \up & 3\%   & 0.78  & \up & 2\%   & 1.68  & \up & 2\% \\
    2-3-4-1 & 0.71  & \up & 11\%  & 4.40  & \up & 14\%  & 3.75  & \up & 7\%   & 0.67  & \down & 14\%  & 0.97  & \down & 41\% \\
    2-4-1-3 & 1.13  & \up & 76\%  & 19.38 & \up & 403\% & 9.68  & \up & 176\% & 1.49  & \up & 403\% & 1.61  & \down & 2\% \\
    2-4-3-1 & 1.23  & \up & 91\%  & 17.47 & \up & 354\% & 9.63  & \up & 174\% & 1.35  & \up & 354\% & 1.61  & \down & 2\% \\ \hline
    3-1-2-4 & 0.72  & \up & 11\%  & 4.53  & \up & 18\%  & 4.07  & \up & 16\%  & 0.74  & \up & 18\%  & 1.68  & \up & 2\% \\
    3-1-4-2 & 1.08  & \up & 68\%  & 8.72  & \up & 126\% & 6.68  & \up & 90\%  & 0.97  & \up & 126\% & 1.62  & \down & 2\% \\
    3-2-1-4 & 0.74  & \up & 15\%  & 3.69  & \down & 4\%   & 3.46  & \up & 1\%   & 0.73  & \up & 4\%   & 1.65  & \up & 1\% \\
    3-2-4-1 & 0.79  & \up & 23\%  & 4.17  & \up & 8\%   & 3.61  & \up & 3\%   & 0.77  & \up & 8\%   & 0.96  & \down & 42\% \\
    3-4-1-2 & 1.01  & \up & 57\%  & 8.79  & \up & 128\% & 6.91  & \up & 97\%  & 1.15  & \up & 128\% & 1.54  & \down & 6\% \\
    3-4-2-1 & 1.11  & \up & 73\%  & 8.79  & \up & 128\% & 7.66  & \up & 118\% & 0.97  & \up & 128\% & 1.40  & \down & 15\% \\ \hline
    4-1-2-3 & 1.18  & \up & 84\%  & 16.39 & \up & 326\% & 9.02  & \up & 157\% & 1.59  & \up & 326\% & 3.20  & \up & 95\% \\
    4-1-3-2 & 1.24  & \up & 92\%  & 6.40  & \up & 66\%  & 6.20  & \up & 77\%  & 1.59  & \up & 66\%  & 3.20  & \up & 95\% \\
    4-2-1-3 & 1.24  & \up & 92\%  & 20.91 & \up & 443\% & 10.01 & \up & 185\% & 1.65  & \up & 443\% & 2.09  & \up & 28\% \\
    4-2-3-1 & 1.35  & \up & 110\% & 19.12 & \up & 396\% & 10.07 & \up & 187\% & 1.53  & \up & 396\% & 2.09  & \up & 28\% \\
    4-3-1-2 & 1.18  & \up & 84\%  & 6.40  & \up & 66\%  & 6.16  & \up & 75\%  & 1.47  & \up & 66\%  & 2.09  & \up & 28\% \\
    4-3-2-1 & 1.31  & \up & 103\% & 8.94  & \up & 132\% & 7.75  & \up & 121\% & 1.25  & \up & 132\% & 2.09  & \up & 28\% \\ \bottomrule    
    \end{tabularx}
    
    \begin{tablenotes} \footnotesize
\item
\emph{Note}:
In columns Chg., the arrow/number shows the direction/size of the change in the non-uniformity of the draw compared to the official draw order 1-2-3-4 with ex-ante group labelling (see the first row).
    \end{tablenotes}
\end{threeparttable}
}
\end{table}

Table~\ref{Table3} shows the non-uniformity of the Skip mechanism with the 24 possible draw orders and ex-ante group labelling for the 2026 FIFA World Cup draw, according to the five measures. Usually, all metrics remain similar if only the first two pots are exchanged. Crucially, the official draw order of Pot 1, Pot 2, Pot 3, Pot 4 performs best with respect to $M_1$ and $M_3$, and the possible improvement is quite limited for measures $M_2$ and $M_4$. The order of the pots plays a non-negligible role: the value of $M_1$ can be doubled by choosing another version of the Skip mechanism, and the situation is even worse for the other three measures.
The maximum difference $M_2$ can exceed 20 percentage points under the draw order 4-2-1-3, when the assignment probability of Paraguay (the only CONMEBOL team in Pot 3) and the winner of the first inter-continental play-off (IC Path 1) equals 18.2\% in a uniform draw but 39.1\% under the Skip mechanism.
However, the official draw procedure is suboptimal with respect to measure $M_5$; it is dominated by 10 draw orders, in some cases by more than 40\%.

According to Table~\ref{Table3}, the average distortion never decreases when Pot 4 comes earlier in a draw order. It is caused by three unique teams in Pot 4, New Zealand and the two winners of the intercontinental play-offs. As we have seen in Section~\ref{Sec33}, the winner of IC Path 2 can play in a group only together with one CAF and two UEFA teams. Regarding the winner of IC Path 1, at most one AFC, at most one CONMEBOL, and at most two UEFA teams can be assigned to its group. Even though no constraint applies directly to New Zealand, it also presents difficulties for the Skip mechanism since each group should contain at least one UEFA team.

The labels of the groups for these three teams are fixed when Pot 4 is drawn; thus, emptying Pot 4 early is usually suboptimal, as shown by Example~\ref{Examp3}. For instance, consider the winner of IC Path 1 in the two extreme cases. When Pot 4 is emptied last, we should have a group with one CAF and two UEFA teams, but there is no constraint on the label of this group. On the other hand, when Pot 4 is emptied first and the winner of IC Path 1 is assigned to group $j$, group $j$ is required to contain one CAF and two UEFA teams, instead of only one of the 12 groups.

According to \citet{Csato2026}, if at most two teams from a given set can be assigned to a group, the optimal draw order contains the groups in a decreasing order with respect to the number of teams from this set. Hence, Pot 1 with its seven UEFA teams should be considered first. Combining this with the above argument to draw Pot 4 last explains the optimality of the official draw order based on average distortion $M_1$.

To support this idea, Table~\ref{Table_A1} in Appendix~\ref{Appendix_B} shows the distortion of the Skip mechanism -- derived from the same number of simulations as before -- when the two winners of intercontinental play-offs are considered only for the confederation with the highest number of teams, CAF (IC Path 1) and AFC (IC Path 2). As it turned out after the draw in March 2026, the winners were indeed Congo DR and Iraq, coming from these confederations. Therefore, Table~\ref{Table_A1} corresponds to a draw conducted once all participating teams are known.
Both the range of average (0.85--1.36 instead of 0.64--1.35) and, especially, maximum (4.27--8.98 instead of 3.85--20.91) biases are smaller in Table~\ref{Table_A1} than in Table~\ref{Table3}. Interestingly, the official draw order 1-2-3-4 becomes more distorted despite the less restrictive set of constraints. Crucially, the draw order 1-2-3-4 is no longer optimal: draw order 3-2-1-4 is better according to all measures $M_1$--$M_5$, and the official draw is even the worst with respect to measure $M_3$.

In order to see why the draw order 4-2-1-3 gives the highest maximum non-uniformity, Table~\ref{Table_A2} in Appendix~\ref{Appendix_B} presents the assignment probabilities for the teams drawn from Pots 3 and 4 under a uniform draw (Table~\ref{Table_A2a}) and the Skip mechanism with the draw order 4-2-1-3 (Table~\ref{Table_A2b}). As we have seen, the maximum bias occurs between Paraguay and the winner of IC Path 1, and can be explained as follows.
The winner of IC Path 1 can play against six opponents from Pot 3, two UEFA teams, three AFC teams, and the only CONMEBOL team Paraguay.
If the draw uses the Skip mechanism and starts with Pot 4, the group of the IC Path 1 winner contains an AFC team from Pot 2 with a probability of about 46.4\%, because AFC teams compose one-third of Pot 2 but they can play only against 10 teams from Pot 4. Then, only the two UEFA teams and Paraguay remain available from Pot 3 to be assigned to the group of the IC Path 1 winner, but there is an extra demand for UEFA teams since each of the 12 groups should contain at least one out of the 16 UEFA teams.
This requirement cannot be neglected. For example, New Zealand is assigned to a group that contains a non-UEFA team from Pot 1 with a probability of roughly 48.3\%. Since four teams in Pot 4 are considered for CONCACAF, New Zealand is assigned to Group A and plays against Mexico with a probability of:
\[
\frac{1}{12} + \frac{4}{12} \cdot \frac{1}{11} + \frac{4}{12} \cdot \frac{3}{11} \cdot \frac{1}{10} + \frac{4}{12} \cdot \frac{3}{11} \cdot \frac{2}{10} \cdot \frac{1}{9} + \frac{4}{12} \cdot \frac{3}{11} \cdot \frac{2}{10} \cdot \frac{1}{9} \cdot \frac{1}{8} = 12.5\%.
\]
This group should contain a UEFA team from either Pot 2 or Pot 3, which have only two UEFA teams each, calling for Norway/Scotland from the last Pot 3 with a substantial probability. Consequently, New Zealand will play against each of Norway and Scotland with a probability of about 17.6\% rather than 8.4\% under a uniform draw.
To conclude, if the Skip mechanism is used with the draw order 4-2-1-3, Paraguay becomes a kind of ``joker'' in Pot 3 to be assigned to a group already containing the winner of IC Path 1 and an Asian team from Pot 2, which occurs in almost half of random draws.

The non-uniformity of the 2026 FIFA World Cup draw seems to be more favourable than in the case of the previous two editions, where the same draw procedure was used.
The value of $M_1$ is about 0.82 and 1.43 for the 2018 and 2022 World Cups, respectively, while the maximum difference $M_2$ reaches 10.24 and 10.49, respectively \citep{Csato2027}. The latter is decreased to roughly one-third of what is seen for eight groups. The reason is partially the 50\% increase in the number of groups that reduces all assignment probabilities by one-third in an unconstrained draw. In addition, having 12 groups instead of eight decreases the role of prohibited pairs imposed by the constraints for all confederations except for UEFA.
To conclude, the expansion to 48 teams with the same structure of draw restrictions has greatly improved the non-uniformity of the group draw, at least for the 2026 FIFA World Cup with its given set of geographic (and competitive balance) restrictions.

%\begin{afterpage}           % This prevents page breaks, but shifts the table.
\begin{knitrout}
\definecolor{shadecolor}{rgb}{0.969, 0.969, 0.969}\color{fgcolor}
\begin{sidewaystable}
\centering\centering
\caption{\label{Table4} The differences of assignment probabilities in the 2026 FIFA World Cup draw for all pairs of national teams}
\centering
\resizebox{\ifdim\width>\linewidth\linewidth\else\width\fi}{!}{
\begin{threeparttable}
\begin{tabular}[t]{lrrrrrrrrrrrrrrrrrrrrrrrrrrrrrrrrrrrr}
\toprule
\rotatebox{90}{ } & \rotatebox{90}{Croatia} & \rotatebox{90}{Morocco} & \rotatebox{90}{Colombia} & \rotatebox{90}{Uruguay} & \rotatebox{90}{Switzerland} & \rotatebox{90}{Japan} & \rotatebox{90}{Senegal} & \rotatebox{90}{Iran} & \rotatebox{90}{South Korea} & \rotatebox{90}{Ecuador} & \rotatebox{90}{Austria} & \rotatebox{90}{Australia} & \rotatebox{90}{Norway} & \rotatebox{90}{Panama} & \rotatebox{90}{Egypt} & \rotatebox{90}{Algeria} & \rotatebox{90}{Scotland} & \rotatebox{90}{Paraguay} & \rotatebox{90}{Tunisia} & \rotatebox{90}{Ivory Coast} & \rotatebox{90}{Uzbekistan} & \rotatebox{90}{Qatar} & \rotatebox{90}{Saudi Arabia} & \rotatebox{90}{South Africa} & \rotatebox{90}{Jordan} & \rotatebox{90}{Cape Verde} & \rotatebox{90}{Ghana} & \rotatebox{90}{Cura{\c c}ao} & \rotatebox{90}{Haiti} & \rotatebox{90}{New Zealand} & \rotatebox{90}{UEFA Path A} & \rotatebox{90}{UEFA Path B} & \rotatebox{90}{UEFA Path C} & \rotatebox{90}{UEFA Path D} & \rotatebox{90}{IC Path 1} & \rotatebox{90}{IC Path 2}\\
\bottomrule
\end{tabular}
\begin{tablenotes}
\item
\emph{Notes}:
Empty cells represent pairs of teams that could not be assigned to the same group.
\item
The numbers show the changes of assignment probabilities in percentage points rounded to two decimal places.
The 99\% confidence interval is the reported value $\pm 0.118$.
\item
Green (Red) colour means that the official draw procedure (Skip mechanism with the draw order 1-2-3-4) implies a higher (lower) probability than a uniform draw. Darker colour indicates a greater change in absolute value.
\item
The three hosts are pre-assigned to a particular group: Mexico to Group A, Canada to Group B, and the United States to Group D.
\item
Teams from the same confederation and seeding pot are interchangeable, except for Argentina, England, France, and Spain. England and France are also interchangeable.
\end{tablenotes}
\end{threeparttable}}
\end{sidewaystable}

\end{knitrout}

%\end{afterpage}

Table~\ref{Table4} presents the differences in assignment probabilities compared to a uniform draw for all pairs of countries if the official draw procedure is used.
The largest values are observed for Mexico and the two African nations drawn from Pot 4: their assignment probability equals 9.1\% in a uniform draw, but remains only 5.3\% in the actual draw. The non-European teams in Pot 1, especially the two South American nations, Argentina and Brazil, are more likely to play against the winners of UEFA play-offs; the chance is about 14\% instead of 11.3\% for each of the four teams. This is caused by the short-sightedness of the Skip mechanism, which considers the upper bound for UEFA nations only after two UEFA teams are assigned to the same group, in contrast to a uniform draw that ``knows'' the presence of four UEFA teams in Pot 4. Since the eight non-UEFA teams in Pot 4 are weaker than the four UEFA teams, the seven strongest European teams benefit from using the Skip mechanism. Argentina and Brazil have a lower probability to play against the five European teams drawn from Pots 2 and 3 in the actual draw.

The three host nations are treated differently due to their pre-assignment to the groups because the Skip mechanism is not independent of the group labels. Mexico -- assigned automatically to Group A -- has a higher chance to play against a strong opponent drawn from Pot 4 ($4 \cdot 19.3\%$) than Canada ($4 \cdot 18\%$), and, especially, the United States ($4 \cdot 17.2\%$).

\begin{table}[t!]
\centering
\caption{Measures of non-uniformity in the 2026 FIFA World Cup draw \\ under all draw orders of the Skip mechanism: ex-post group labelling}
\label{Table5}

\centerline{
\begin{threeparttable}
\rowcolors{3}{}{gray!20}
    \begin{tabularx}{1.2\linewidth}{l Rrr Rrr Rrr Rrr Rrr} \toprule \hiderowcolors
    Measure & \multicolumn{3}{c}{Average ($M_1$)} & \multicolumn{3}{c}{Maximum ($M_2$)} & \multicolumn{3}{c}{Maximum 8 ($M_3$)} & \multicolumn{3}{c}{Pot 4 ($M_4$)} & \multicolumn{3}{c}{Pots 1/4 ($M_5$)} \\
    Order & Value  & \multicolumn{2}{c}{Chg.} & Value  & \multicolumn{2}{c}{Chg.} & Value  & \multicolumn{2}{c}{Chg.} & Value  & \multicolumn{2}{c}{Chg.} & Value  & \multicolumn{2}{c}{Chg.} \\ \bottomrule \showrowcolors
    1-2-3-4 & 0.76  & \up & 19\%  & 3.44  & \down & 11\%  & 3.15  & \down & 10\%  & 0.78  & \up & 12\%  & 1.35  & \down & 18\% \\
    1-2-4-3 & 0.96  & \up & 50\%  & 15.51 & \up & 303\% & 6.94  & \up & 98\%  & 1.02  & \up & 47\%  & 1.23  & \down & 25\% \\
    1-3-2-4 & 0.78  & \up & 22\%  & 4.53  & \up & 18\%  & 3.90  & \up & 11\%  & 0.87  & \up & 25\%  & 1.49  & \down & 9\% \\
    1-3-4-2 & 1.04  & \up & 61\%  & 6.13  & \up & 59\%  & 5.55  & \up & 58\%  & 1.05  & \up & 51\%  & 1.82  & \up & 11\% \\
    1-4-2-3 & 1.15  & \up & 78\%  & 16.13 & \up & 319\% & 7.96  & \up & 127\% & 1.65  & \up & 138\% & 3.20  & \up & 95\% \\
    1-4-3-2 & 1.11  & \up & 73\%  & 6.31  & \up & 64\%  & 5.96  & \up & 70\%  & 1.62  & \up & 133\% & 3.20  & \up & 95\% \\ \hline
    2-1-3-4 & 0.77  & \up & 20\%  & 3.41  & \down & 12\%  & 3.13  & \down & 11\%  & 0.78  & \up & 12\%  & 1.31  & \down & 20\% \\
    2-1-4-3 & 0.97  & \up & 51\%  & 15.53 & \up & 303\% & 6.99  & \up & 99\%  & 1.04  & \up & 49\%  & 1.25  & \down & 24\% \\
    2-3-1-4 & 0.91  & \up & 41\%  & 3.66  & \down & 5\%   & 3.60  & \up & 2\%   & 0.85  & \up & 22\%  & 1.34  & \down & 18\% \\
    2-3-4-1 & 0.85  & \up & 33\%  & 4.93  & \up & 28\%  & 4.41  & \up & 25\%  & 0.69  & \down & 0\%   & 0.79  & \down & 52\% \\
    2-4-1-3 & 1.05  & \up & 64\%  & 17.10 & \up & 344\% & 9.04  & \up & 157\% & 1.38  & \up & 99\%  & 0.82  & \down & 50\% \\
    2-4-3-1 & 1.03  & \up & 60\%  & 8.70  & \up & 126\% & 7.11  & \up & 103\% & 1.18  & \up & 70\%  & 1.17  & \down & 29\% \\ \hline
    3-1-2-4 & 0.79  & \up & 23\%  & 4.57  & \up & 19\%  & 3.88  & \up & 11\%  & 0.86  & \up & 24\%  & 1.46  & \down & 11\% \\
    3-1-4-2 & 1.04  & \up & 62\%  & 6.23  & \up & 62\%  & 5.60  & \up & 59\%  & 1.06  & \up & 53\%  & 1.82  & \up & 11\% \\
    3-2-1-4 & 0.91  & \up & 42\%  & 3.67  & \down & 5\%   & 3.62  & \up & 3\%   & 0.87  & \up & 25\%  & 1.36  & \down & 17\% \\
    3-2-4-1 & 0.84  & \up & 30\%  & 4.71  & \up & 22\%  & 4.33  & \up & 23\%  & 0.69  & \down & 1\%   & 0.83  & \down & 49\% \\
    3-4-1-2 & 1.00  & \up & 55\%  & 5.98  & \up & 55\%  & 5.06  & \up & 44\%  & 1.38  & \up & 99\%  & 1.79  & \up & 9\% \\
    3-4-2-1 & 0.82  & \up & 28\%  & 3.37  & \down & 12\%  & 3.17  & \down & 10\%  & 1.13  & \up & 63\%  & 1.14  & \down & 31\% \\ \hline
    4-1-2-3 & 1.15  & \up & 79\%  & 15.64 & \up & 306\% & 7.87  & \up & 124\% & 1.66  & \up & 139\% & 3.20  & \up & 95\% \\
    4-1-3-2 & 1.13  & \up & 75\%  & 6.30  & \up & 63\%  & 5.99  & \up & 71\%  & 1.62  & \up & 133\% & 3.20  & \up & 95\% \\
    4-2-1-3 & 1.04  & \up & 62\%  & 16.94 & \up & 340\% & 8.94  & \up & 155\% & 1.35  & \up & 94\%  & 0.74  & \down & 55\% \\
    4-2-3-1 & 1.02  & \up & 59\%  & 8.61  & \up & 124\% & 7.03  & \up & 100\% & 1.20  & \up & 73\%  & 1.25  & \down & 24\% \\
    4-3-1-2 & 0.99  & \up & 54\%  & 5.92  & \up & 54\%  & 5.01  & \up & 43\%  & 1.37  & \up & 97\%  & 1.70  & \up & 3\% \\
    4-3-2-1 & 0.83  & \up & 30\%  & 3.50  & \down & 9\%   & 3.25  & \down & 7\%   & 1.16  & \up & 67\%  & 1.26  & \down & 23\% \\
 \bottomrule    
    \end{tabularx}
    
    \begin{tablenotes} \footnotesize
\item
\emph{Note}:
In columns Chg., the arrow/number shows the direction/size of the change in the non-uniformity of the draw compared to the official draw order 1-2-3-4 with ex-ante group labelling (see the first row in Table~\ref{Table3}).
    \end{tablenotes}
\end{threeparttable}
}
\end{table}

This source of non-uniformity can be avoided by labelling the groups only after their composition is determined. 
Thus, Table~\ref{Table5} reports the results with ex-post group labelling. Unsurprisingly, the highest difference(s) in assignment probabilities is (are) lower compared to the official draw (see $M_2$ and $M_3$) as Mexico, Canada, and the United States have only 17.4\% chance to play against a given winner of the UEFA play-offs. On the other hand, the value of $M_1$ increases by almost 20\%, and ex-post group labelling is not effective under any draw order with respect to measure $M_1$. The maximum reduction of about 10\% in $M_2$ and $M_3$  probably does not compensate for the worsening average difference in all assignment probabilities.

However, the original value of $M_4$ can be retained, and $M_5$ -- the mean difference in the assignment probability of a team in Pot 1 and a UEFA team in Pot 4 -- can be more than halved by choosing the draw order 2-3-4-1 or 3-2-4-1. Ex-post group labelling is worth using with one of these draw orders if the organiser is willing to accept a decline of 25--30\% in measures $M_1$--$M_3$ in order to mitigate the impact of the imbalanced Pot 4 on the seeded teams in Pot 1.
Even though draw orders 2-4-3-1 and 4-2-3-1 also lead to a comparable reduction in $M_5$, they greatly increase $M_3$ and, especially, $M_2$, by distorting at least eight assignment probabilities by 9--17 percentage points compared to a uniform draw.

\section{Policy implications and conclusions} \label{Sec5}

The organisers of sports tournaments appear to pay limited attention to the non-uniformity of the group draw, despite its non-negligible sporting effects \citep{Csato2025c}. This is suboptimal since the Skip mechanism, which is currently the most popular procedure used to enforce various constraints in a group draw \citep{Csato2025f}, has several variants with possibly different levels of non-uniformity. In the case of the 2026 FIFA World Cup draw, the decision-makers can choose from 48 reasonable designs -- depending on the order of the four pots and the group labelling policy -- that require only marginal modifications in the draw procedure.

The framework proposed above allows finding an optimal mechanism according to the given objective function. In particular, while the official draw procedure can barely be improved with respect to most measures of non-uniformity, it systematically favours the UEFA teams and punishes the non-UEFA teams drawn from Pot 1, including the three hosts. The degree of non-uniformity can be reduced by more than 50\% using a different order of the pots and switching to ex-post group labelling.

As it has turned out during our research, the 2026 FIFA World Cup draw is substantially more difficult to simulate than the 2018 and 2022 FIFA World Cup draws due to the higher number of teams and the complexity of the draw restrictions. Hence, the previously suggested recursive backtracking algorithms become intractable even on a supercomputer. We have developed a novel implementation of the Skip mechanism via integer programming to solve this problem.

These challenges suggest that a stronger collaboration between policy-makers and the academic community is indispensable to optimise the rules of group draws in sports tournaments. Hopefully, the current paper will inspire future projects with this aim.

\section*{Acknowledgements}
\addcontentsline{toc}{section}{Acknowledgements}

\emph{Julien Guyon} and two anonymous reviewers provided valuable comments and suggestions on an earlier draft. \\
The research was supported by the National Research, Development and Innovation Office under Grants Advanced 152220 and FK 145838, the J\'anos Bolyai Research Scholarship of the Hungarian Academy of Sciences, and the Special Research Fund [BOF/24J/2021/188] of Ghent University.
%Gurobi \citep{Gurobi} was used to solve the LPs; R with the additional packages doRNG and kableExtra \citep{R2025, doRNG, kableExtra} was used to implement the uniform sampler and to prepare some of the results. 

\bibliographystyle{apalike} 
\bibliography{All_references}

\clearpage
\renewcommand{\thesubsection}{\Alph{subsection}}

\section*{Appendix}
\addcontentsline{toc}{section}{Appendix}

\setcounter{table}{0}
\renewcommand{\thetable}{A.\arabic{table}}

\subsection{Note on balanced distribution of the highest-ranked teams}  \label{Appendix_A}

The official description of the draw procedure for the 2026 FIFA World Cup \citep{FIFA2025b} is difficult to understand with respect to the draw constraint related to competitive balance: \\
``\emph{In the interest of ensuring competitive balance, two separate pathways to the semi-finals have been established, as per Appendix A below. In accordance with this and in order to have a balanced distribution of the teams, the four highest-ranked teams in the FIFA/Coca-Cola Men's World Ranking, when drawn, will have the following constraints. The highest-ranked team and the second highest-ranked team will be randomly drawn into groups in opposite pathways, and the same principle will also apply to the third and fourth highest-ranked teams. This will ensure that, should they win their groups, the two highest-ranked teams would not meet before the final.}'' \\
The issue is clarified by the video of the 2026 FIFA World Cup
draw (\url{https://www.youtube.com/watch?v=9HX_tQBA-Iw}), approximately from 1:25:43 to 1:26:17: \\
``\emph{The top four teams in the FIFA/Coca-Cola World Ranking will be drawn in a way where the top two, Spain and Argentina, must be drawn in different pathways. So that means if they both win their group, they would only meet in the final if they make it that far. Now the third- and fourth-placed teams, France and England, also needs to be drawn in different pathways, so that if they win their group, they would only meet the top two ranked teams in the semifinals. This is only on the basis that all four teams win their group and get that far.}'' \\
This is exactly how we described the constraint in Section~\ref{Sec31} and formulated it in Section~\ref{Sec34}.

\clearpage
\subsection{Supplementary tables}  \label{Appendix_B}

\begin{table}[ht!]
\centering
\caption{Measures of non-uniformity in the modified 2026 FIFA World Cup draw: ex-ante group labelling, IC Path winners are assigned to one confederation only}
\label{Table_A1}

\centerline{
\begin{threeparttable}
\rowcolors{3}{}{gray!20}
    \begin{tabularx}{1.2\linewidth}{l Rrr Rrr Rrr Rrr Rrr} \toprule \hiderowcolors
    Measure & \multicolumn{3}{c}{Average ($M_1$)} & \multicolumn{3}{c}{Maximum ($M_2$)} & \multicolumn{3}{c}{Maximum 8 ($M_3$)} & \multicolumn{3}{c}{Pot 4 ($M_4$)} & \multicolumn{3}{c}{Pots 1/4 ($M_5$)} \\
    Order & Value  & \multicolumn{2}{c}{Chg.} & Value  & \multicolumn{2}{c}{Chg.} & Value  & \multicolumn{2}{c}{Chg.} & Value  & \multicolumn{2}{c}{Chg.} & Value  & \multicolumn{2}{c}{Chg.} \\ \bottomrule \showrowcolors
    1-2-3-4 & 0.95  & \multicolumn{2}{c}{---} & 4.49  & \multicolumn{2}{c}{---} & 1.16  & \multicolumn{2}{c}{---} & 4.64  & \multicolumn{2}{c}{---} & 2.66  & \multicolumn{2}{c}{---} \\
    1-2-4-3 & 1.09  & \up & 15\%  & 7.75  & \up & 73\%  & 0.78  & \down & 33\%  & 8.76  & \up & 73\%  & 1.16  & \down & 56\% \\
    1-3-2-4 & 0.91  & \down & 4\%   & 4.27  & \down & 5\%   & 1.14  & \down & 2\%   & 4.46  & \down & 5\%   & 2.52  & \down & 5\% \\
    1-3-4-2 & 1.08  & \up & 14\%  & 5.96  & \up & 33\%  & 0.93  & \down & 20\%  & 7.57  & \up & 33\%  & 1.81  & \down & 32\% \\
    1-4-2-3 & 1.19  & \up & 26\%  & 7.73  & \up & 72\%  & 1.52  & \down & 31\%  & 9.71  & \up & 72\%  & 3.06  & \up & 15\% \\
    1-4-3-2 & 1.15  & \up & 22\%  & 5.92  & \up & 32\%  & 1.50  & \down & 29\%  & 6.78  & \up & 32\%  & 3.06  & \up & 15\% \\ \hline
    2-1-3-4 & 0.93  & \down & 2\%   & 4.51  & \up & 1\%   & 1.13  & \down & 3\%   & 4.69  & \up & 1\%   & 2.63  & \down & 1\% \\
    2-1-4-3 & 1.08  & \up & 14\%  & 7.82  & \up & 74\%  & 0.79  & \down & 32\%  & 8.87  & \up & 74\%  & 1.18  & \down & 55\% \\
    2-3-1-4 & 0.85  & \down & 11\%  & 4.36  & \down & 3\%   & 1.02  & \down & 12\%  & 4.76  & \up & 3\%   & 2.32  & \down & 13\% \\
    2-3-4-1 & 0.85  & \down & 11\%  & 4.60  & \up & 2\%   & 1.01  & \down & 13\%  & 5.08  & \up & 2\%   & 1.91  & \down & 28\% \\
    2-4-1-3 & 1.19  & \up & 25\%  & 8.37  & \up & 86\%  & 1.41  & \down & 21\%  & 9.32  & \up & 86\%  & 1.39  & \down & 48\% \\
    2-4-3-1 & 1.27  & \up & 34\%  & 8.90  & \up & 98\%  & 1.28  & \down & 10\%  & 11.23 & \up & 98\%  & 1.39  & \down & 48\% \\ \hline
    3-1-2-4 & 0.91  & \down & 4\%   & 4.28  & \down & 5\%   & 1.13  & \down & 3\%   & 4.49  & \down & 5\%   & 2.51  & \down & 5\% \\
    3-1-4-2 & 1.07  & \up & 13\%  & 5.96  & \up & 33\%  & 0.91  & \down & 22\%  & 7.48  & \up & 33\%  & 1.78  & \down & 33\% \\
    3-2-1-4 & 0.87  & \down & 9\%   & 4.27  & \down & 5\%   & 1.06  & \down & 9\%   & 4.56  & \down & 5\%   & 2.29  & \down & 14\% \\
    3-2-4-1 & 0.89  & \down & 6\%   & 4.59  & \up & 2\%   & 1.09  & \down & 7\%   & 4.84  & \up & 2\%   & 1.85  & \down & 31\% \\
    3-4-1-2 & 1.06  & \up & 11\%  & 6.11  & \up & 36\%  & 1.10  & \down & 6\%   & 7.55  & \up & 36\%  & 1.54  & \down & 42\% \\
    3-4-2-1 & 1.03  & \up & 9\%   & 6.94  & \up & 55\%  & 0.89  & \down & 24\%  & 7.54  & \up & 55\%  & 1.20  & \down & 55\% \\ \hline
    4-1-2-3 & 1.19  & \up & 26\%  & 7.72  & \up & 72\%  & 1.52  & \down & 31\%  & 9.73  & \up & 72\%  & 3.06  & \up & 15\% \\
    4-1-3-2 & 1.15  & \up & 22\%  & 5.93  & \up & 32\%  & 1.50  & \down & 29\%  & 6.71  & \up & 32\%  & 3.05  & \up & 15\% \\
    4-2-1-3 & 1.24  & \up & 31\%  & 8.29  & \up & 85\%  & 1.62  & \down & 39\%  & 10.04 & \up & 85\%  & 1.94  & \down & 27\% \\
    4-2-3-1 & 1.36  & \up & 43\%  & 8.98  & \up & 100\% & 1.50  & \down & 29\%  & 12.10 & \up & 100\% & 1.94  & \down & 27\% \\
    4-3-1-2 & 1.14  & \up & 20\%  & 6.07  & \up & 35\%  & 1.42  & \down & 22\%  & 7.03  & \up & 35\%  & 2.09  & \down & 21\% \\
    4-3-2-1 & 1.17  & \up & 23\%  & 7.09  & \up & 58\%  & 1.16  & \down & 0\%   & 8.35  & \up & 58\%  & 1.94  & \down & 27\% \\ \bottomrule    
    \end{tabularx}
    
    \begin{tablenotes} \footnotesize
\item
\emph{Notes}:
In columns Chg., the arrow/number shows the direction/size of the change in the non-uniformity of the draw compared to the official draw order 1-2-3-4 with ex-ante group labelling (see the first row).
\item
The winner of IC Path 1 is prohibited to play against a CAF nation, but it can play against a CONCACAF or OFC team. Indeed, the winner was Congo DR from CAF.
\item
The winner of IC Path 2 is prohibited to play against an AFC nation, but it can play against a CONCACAF or CONMEBOL team. Indeed, the winner was Iraq from AFC.
    \end{tablenotes}
\end{threeparttable}
}
\end{table}

\begin{table}[t!]
\centering
\caption{Assignment probabilities between \\ Pots 3 and 4 in the 2026 FIFA World Cup draw}
\label{Table_A2}

\begin{subtable}{\textwidth}
\subcaption{Uniform draw}
\label{Table_A2a}

\centerline{
\rowcolors{3}{}{gray!20}
    \begin{tabularx}{1.2\linewidth}{l RRR RRR RRR RrR} \toprule \hiderowcolors
          & \rotatebox{90}{Jordan} & \rotatebox{90}{Cabo Verde} & \rotatebox{90}{Ghana} & \rotatebox{90}{Cura{\c c}ao} & \rotatebox{90}{Haiti} & \rotatebox{90}{New Zealand} & \rotatebox{90}{UEFA Path A} & \rotatebox{90}{UEFA Path B} & \rotatebox{90}{UEFA Path C} & \rotatebox{90}{UEFA Path D} & \rotatebox{90}{IC Path 1} & \rotatebox{90}{IC Path 2} \\ \bottomrule \showrowcolors
    Norway & 10.68 & 15.29 & 15.29 & 5.89  & 5.89  & 8.40  & 2.80  & 2.80  & 2.80  & 2.79  & 14.00 & 13.36 \\
    Panama & 14.88 & 21.34 & 21.35 &       &       & 10.74 & 7.93  & 7.93  & 7.92  & 7.92  &       &  \\
    Egypt & 11.12 &       &       & 10.88 & 10.89 & 8.83  & 10.91 & 10.90 & 10.90 & 10.91 &       & 14.66 \\
    Algeria & 11.12 &       &       & 10.89 & 10.89 & 8.83  & 10.90 & 10.90 & 10.90 & 10.91 &       & 14.66 \\
    Scotland & 10.68 & 15.29 & 15.30 & 5.90  & 5.89  & 8.40  & 2.79  & 2.80  & 2.79  & 2.79  & 13.99 & 13.37 \\
    Paraguay & 8.17  & 11.61 & 11.61 & 8.03  & 8.03  & 6.63  & 6.92  & 6.92  & 6.92  & 6.92  & \textbf{18.22} &  \\
    Tunisia & 11.12 &       &       & 10.89 & 10.89 & 8.83  & 10.91 & 10.90 & 10.90 & 10.90 &       & 14.66 \\
    Ivory Coast & 11.11 &       &       & 10.88 & 10.89 & 8.84  & 10.91 & 10.91 & 10.90 & 10.91 &       & 14.65 \\
    Uzbekistan &       & 12.16 & 12.15 & 8.59  & 8.58  & 7.22  & 8.34  & 8.35  & 8.35  & 8.34  & 17.93 &  \\
    Qatar &       & 12.15 & 12.15 & 8.58  & 8.58  & 7.22  & 8.34  & 8.35  & 8.35  & 8.35  & 17.93 &  \\
    Saudi Arabia &       & 12.15 & 12.15 & 8.58  & 8.57  & 7.22  & 8.35  & 8.35  & 8.35  & 8.35  & 17.94 &  \\
    South Africa & 11.12 &       &       & 10.88 & 10.89 & 8.83  & 10.90 & 10.91 & 10.91 & 10.90 &       & 14.65 \\ \bottomrule    
    \end{tabularx}
}
\end{subtable}

\vspace{0.5cm}
\begin{subtable}{\textwidth}
\subcaption{Skip mechanism with the draw order 4-2-1-3}
\label{Table_A2b}

\centerline{
\begin{threeparttable}
\rowcolors{3}{}{gray!20}
    \begin{tabularx}{1.2\linewidth}{l RRR RRR RRR RrR} \toprule \hiderowcolors
          & \rotatebox{90}{Jordan} & \rotatebox{90}{Cabo Verde} & \rotatebox{90}{Ghana} & \rotatebox{90}{Cura{\c c}ao} & \rotatebox{90}{Haiti} & \rotatebox{90}{New Zealand} & \rotatebox{90}{UEFA Path A} & \rotatebox{90}{UEFA Path B} & \rotatebox{90}{UEFA Path C} & \rotatebox{90}{UEFA Path D} & \rotatebox{90}{IC Path 1} & \rotatebox{90}{IC Path 2} \\ \bottomrule \showrowcolors
    Norway & 14.13 & 16.78 & 16.74 & 4.46  & 4.47  & 17.53 & 0.71  & 0.71  & 0.70  & 0.70  & 5.57  & 17.52 \\
    Panama & 15.16 & 23.25 & 23.11 &       &       & 7.20  & 7.81  & 7.83  & 7.83  & 7.81  &       &  \\
    Egypt & 9.97  &       &       & 11.54 & 11.48 & 7.31  & 11.68 & 11.66 & 11.69 & 11.69 &       & 13.00 \\
    Algeria & 9.98  &       &       & 11.51 & 11.49 & 7.27  & 11.66 & 11.65 & 11.72 & 11.71 &       & 13.00 \\
    Scotland & 14.10 & 16.72 & 16.78 & 4.43  & 4.49  & 17.61 & 0.69  & 0.69  & 0.69  & 0.70  & 5.59  & 17.54 \\
    Paraguay & 6.69  & 10.58 & 10.61 & 5.74  & 5.75  & 3.87  & 4.40  & 4.42  & 4.41  & 4.39  & \textbf{39.12} &  \\
    Tunisia & 10.05 &       &       & 11.50 & 11.53 & 7.25  & 11.71 & 11.66 & 11.70 & 11.63 &       & 12.97 \\
    Ivory Coast & 9.94  &       &       & 11.49 & 11.51 & 7.31  & 11.68 & 11.68 & 11.70 & 11.65 &       & 13.03 \\
    Uzbekistan &       & 10.90 & 10.90 & 9.28  & 9.27  & 5.79  & 9.33  & 9.35  & 9.29  & 9.33  & 16.56 &  \\
    Qatar &       & 10.87 & 10.96 & 9.28  & 9.29  & 5.80  & 9.27  & 9.32  & 9.32  & 9.35  & 16.54 &  \\
    Saudi Arabia &       & 10.90 & 10.91 & 9.27  & 9.24  & 5.75  & 9.34  & 9.32  & 9.31  & 9.35  & 16.61 &  \\
    South Africa & 9.99  &       &       & 11.51 & 11.48 & 7.30  & 11.71 & 11.71 & 11.67 & 11.70 &       & 12.94 \\ \bottomrule    
    \end{tabularx}
    
    \begin{tablenotes} \footnotesize
\item
\emph{Notes}:
Empty cells represent pairs of teams that could not be assigned to the same group.
\item
The numbers show the assignment probabilities in percentage rounded to two decimal places.
\item
Teams from the same confederation and seeding pot are interchangeable.
    \end{tablenotes}
\end{threeparttable}
}
\end{subtable}

\end{table}

\end{document}